\documentclass{aa}
\usepackage{txfonts}
\usepackage{graphicx}
\usepackage{natbib}

\bibpunct{(}{)}{;}{a}{}{,}

\usepackage{color}

\newcommand{\kms}{\,{\rm km\,s}^{-1}}

\newcommand{\Msun}{\,\mathrm{M}_\odot}

\newcommand{\as}{\ifmmode {^{\scriptscriptstyle\prime\prime}}
        \else $^{\scriptscriptstyle\prime\prime}$\fi}

\newcommand{\simless}{\mathbin{\lower 3pt\hbox
      {$\rlap{\raise 5pt\hbox{$\char'074$}}\mathchar"7218$}}} 
\newcommand{\simgreat}{\mathbin{\lower 3pt\hbox
     {$\rlap{\raise 5pt\hbox{$\char'076$}}\mathchar"7218$}}} 

\renewcommand{\textbf}[1]{\rm #1}

\newcommand{\rout}{185\,\mathrm{au}}
\newcommand{\xpco}{X_p(\mathrm{CO})}
\newcommand{\xuco}{X_u(\mathrm{CO})}

\begin{document}

\title{The Flying Saucer: \\
Tomography of the thermal and density gas structure  of an edge-on protoplanetary disk.
%
}
\author{
A. Dutrey\inst{1}, S. Guilloteau \inst{1}, V. Pi\'etu \inst{2}, E. 
Chapillon\inst{1,2}, V. Wakelam\inst{1}, E. Di Folco \inst{1}, 
T. Stoecklin\inst{3}, O. Denis-Alpizar\inst{4}, U. Gorti\inst{5}, 
R. Teague\inst{6}, T. Henning\inst{6}, D. Semenov\inst{6},  and
N. Grosso\inst{7}
}
\institute{
Laboratoire d'astrophysique de Bordeaux, Univ. Bordeaux, CNRS, 
B18N, allée Geoffroy Saint-Hilaire, 33615 Pessac, France
  \email{anne.dutrey@u-bordeaux.fr}
\and
IRAM, 300 rue de la piscine, F-38406
Saint Martin d'H\`eres, France
\and
Institut des Sciences Mol\'eculaires, UMR5255-CNRS, 
351 Cours de la libération, F-33405 Talence France
\and
Universidad Autónoma de Chile, 
El llano subercaseaux 2801
San Miguel, Santiago de Chile
\and
SETI Institute /  NASA Ames Research Center, Mail Stop 245-3, Moffett Field, CA 94035-1000, USA
\and
Max-Planck-Institute f\"ur Astronomie, K\"onigstuhl 17, D-69117 Heidelberg, Germany
\and
Observatoire Astronomique de Strasbourg, Universit\'e de Strasbourg, CNRS, UMR 7550, 11 rue de l'Universit\'e, 67000 Strasbourg, France
}

\offprints{A.Dutrey, \email{anne.dutrey@u-bordeaux.fr}}

\date{2017 / 2017} %
\authorrunning{Dutrey et al.} %
\titlerunning{Radial and vertical structure of the Flying Saucer}

\abstract
{Determining the gas density and temperature structures of protoplanetary disks is a 
fundamental task to constrain planet formation theories. This is a challenging 
procedure and most determinations are based on model-dependent assumptions. }
{We attempt a direct determination of the radial and vertical temperature structure of
the Flying Saucer disk, thanks to its favorable inclination of 90 degrees.}
{We present a method based on the tomographic study of an edge-on disk. 
Using ALMA, we observe at 0.5$''$ resolution the Flying Saucer in  CO 
J=2-1 and CS J=5-4. This edge-on disk appears in silhouette against the 
CO J=2-1 emission from background molecular clouds in $\rho$ Oph. The 
combination of velocity gradients due to the Keplerian rotation of the 
disk and intensity variations in the CO background as a function of 
velocity provide a direct measure of the gas temperature as a function 
of radius and height above the disk mid-plane.}
{The overall thermal structure is consistent with model predictions, 
with a cold ($< 15-12 $~K), CO-depleted mid-plane, and a warmer disk 
atmosphere. However, we find evidence for CO gas along the mid-plane 
beyond a radius of about 200\,au, coincident with a change of grain 
properties. Such a behavior is expected in case of efficient rise of UV 
penetration re-heating the disk and thus allowing CO thermal desorption 
or favoring direct CO photo-desorption. CO is also detected up to 3-4 
scale heights while CS is confined around 1 scale height above the 
mid-plane. The limits of the method due to finite spatial and spectral 
resolutions are also discussed.} 
{This method appears to be very promising to determine the gas structure of
planet-forming disks, provided that the molecular data have an angular 
resolution which is high enough, of the order of $0.3 - 0.1''$ at the distance of 
the nearest star forming regions. }


\maketitle{}

\section{Introduction}

Protoplanetary disks orbiting young pre-main sequence stars are the 
sites of planetary system formation. In these disks, gas represents 
about 99$\%$ of the mass and is mostly in the form of H$_2$. 
Since a 
minimum mass of 0.01 $\Msun$ has been determined by  
\citet{Weidenschilling+1977} for the Proto-solar Nebula based on 
the current Solar System, models of planetary systems formation have drastically 
evolved. 
Observational determinations of vertical and radial mass distribution 
of protoplanetary disks provide key constraints for planet formation 
models.  Furthermore, studying the 
gas and dust distributions in protoplanetary disks found around low-mass 
T\,Tauri stars, recognized as young analogs to the Solar System, has 
become a major challenge to understand how planetary systems form 
and evolve.  

With the advent of ALMA, many new results such as the observation of 
narrow dust rings in the HL Tau dust disk \citep{Alma+etal_2015} are 
changing our views on these objects. A series of studies of the disk associated 
with TW Hydrae - the closest T Tauri star - has significantly  
improved our understanding of disk physics and chemistry. This disk is seen 
almost face-on, maximizing its surface, and the dust and gas 
distributions have been intensively observed and modeled 
\citep[]{Qi+etal_2004,Andrews+etal_2012,Rosenfeld+etal_2012,Qi+etal_2013,Andrews+etal_2016,van_Boekel+etal_2017,Teague+etal_2017}.
The J=1-0 transition of HD, the 
hydrogen deuteride,  has been also detected by \citet{Bergin+etal_2013} 
who determined the gas  mass of the disk to be $> 0.056 \Msun$,  a 
value ranging at the upper end of previous estimates based on indirect 
mass tracers ( 5$\cdot 10^{-4}$ - 0.06 $\Msun$, \citep[]{Thi+etal_2010, 
Gorti+etal_2011}. More recently \citet{Teague+etal_2016}  used simple 
molecules such as CO, CN or CS to determine the turbulence inside the 
disk and found that the turbulent line broadening is less than $0.05\,\kms$.
\citet{Schwarz+etal_2016} re-analyzed the HD observations and 
confirmed the high mass of the disk. However, both studies
remain limited by the knowledge of the disk vertical structure, in particular 
the thermal profile of the gas, and have to make assumptions
on the vertical location of molecules, since this cannot be
directly recovered in a face-on disk.

Contrary to a face-on object, the disk around HD\,163296, a Herbig Ae 
star of 2$\Msun$, is inclined by about 45$^\circ$ along the line of sight. 
Such an inclination is enough to partially reveal the vertical location 
of the molecular layer, confirming that a significant fraction of the 
mid-plane is devoid of CO emission \citep{Gregorio-Monsalvo+etal_2013, 
Rosenfeld+etal_2013}. In the case of IM Lupi, a 1$\Msun$ star 
surrounded by a disk inclined by about 45$^\circ$, a multi-line CO analysis 
coupled to a study of the dust disk (images and SED) allowed 
\citet{Cleeves+etal_2016} to provide a coherent picture of the gas and 
dust disk. However, due to the combination of Keplerian shear and 
inclination, a complete determination of the vertical structure is 
challenging because at a given velocity can correspond to several 
locations (radii) inside the disk \citep[see][]{Beckwith+Sargent_1993}. 
In other words, there is no perfect correspondence between a radius and a 
velocity and this generates degeneracies, which are particularly important
when the spatial resolution is limited. A  
purely edge-on disk can allow the retrieval of the full vertical 
structure of the molecules from which the density and temperature 
vertical gradients can be derived, provided the angular resolution is 
high enough and the molecular transitions are adequately selected.   

To test the ability of deriving the disk vertical structure from an 
edge-on disk, we submitted the ALMA project 2013.1.00387.S  
dedicated to the study of the Flying Saucer.  
The \object{Flying Saucer} (\object{2MASS J16281370-2431391})
is an isolated, edge-on disk in the outskirts
of the $\rho$ Oph dark cloud L\,1688 \citep{Grosso+etal_2003} with
evidence for 5-10 $\mu$m-sized dust grains in the upper layers \citep{Pontoppidan+etal_2007}.
\citet{Grosso+etal_2003} resolved  the light scattered by micron-sized dust grains
in near-infrared with the NTT and the VLT and estimated from the nebula extension
dust a disk 
radius of $2.15''$, which is about 260 au for the adopted distance of 
120 pc \citep{Loinard+etal_2008}. The detection of the CN N=2-1 line 
\citep{Reboussin+etal_2015} indicated the existence of a large gas 
disk. The $\rho$ Oph region is crowded with molecular clouds that are 
are strongly emitting in CO lines. The low extinction derived by 
\citet{Grosso+etal_2003} toward the Flying Saucer suggests it lies in 
front of these clouds, and this is confirmed by the CO study of 
\citet{Guilloteau+etal_2016a}. 

We observed CO J=2-1, CS J=5-4, CN N=2-1. 
This is a set of 
standard lines which has been extensively used to retrieve disk 
structures \citep[]{Dartois+etal_2003, Pietu+etal_2007, 
Chapillon+etal_2012, Rosenfeld+etal_2013}.

The dust and CO emissions detected in this ALMA project were partly 
discussed in \citet{Guilloteau+etal_2016a} where we analyzed the 
absorption of the CO background cloud by the dust disk, deriving a dust 
temperature of about 7 K in the dust disk mid-plane at 100 au. This second 
paper deals with the retrieval of the gas temperature 
and density structures based on the analysis of the low angular 
resolution CO and CS lines. After showing the results and the analysis 
of the data, we then discuss the ability of using edge-on disks to 
determine the vertical structure of gas disks.  


\section{Observations}

\begin{figure*}
     \includegraphics[height=11.65cm]{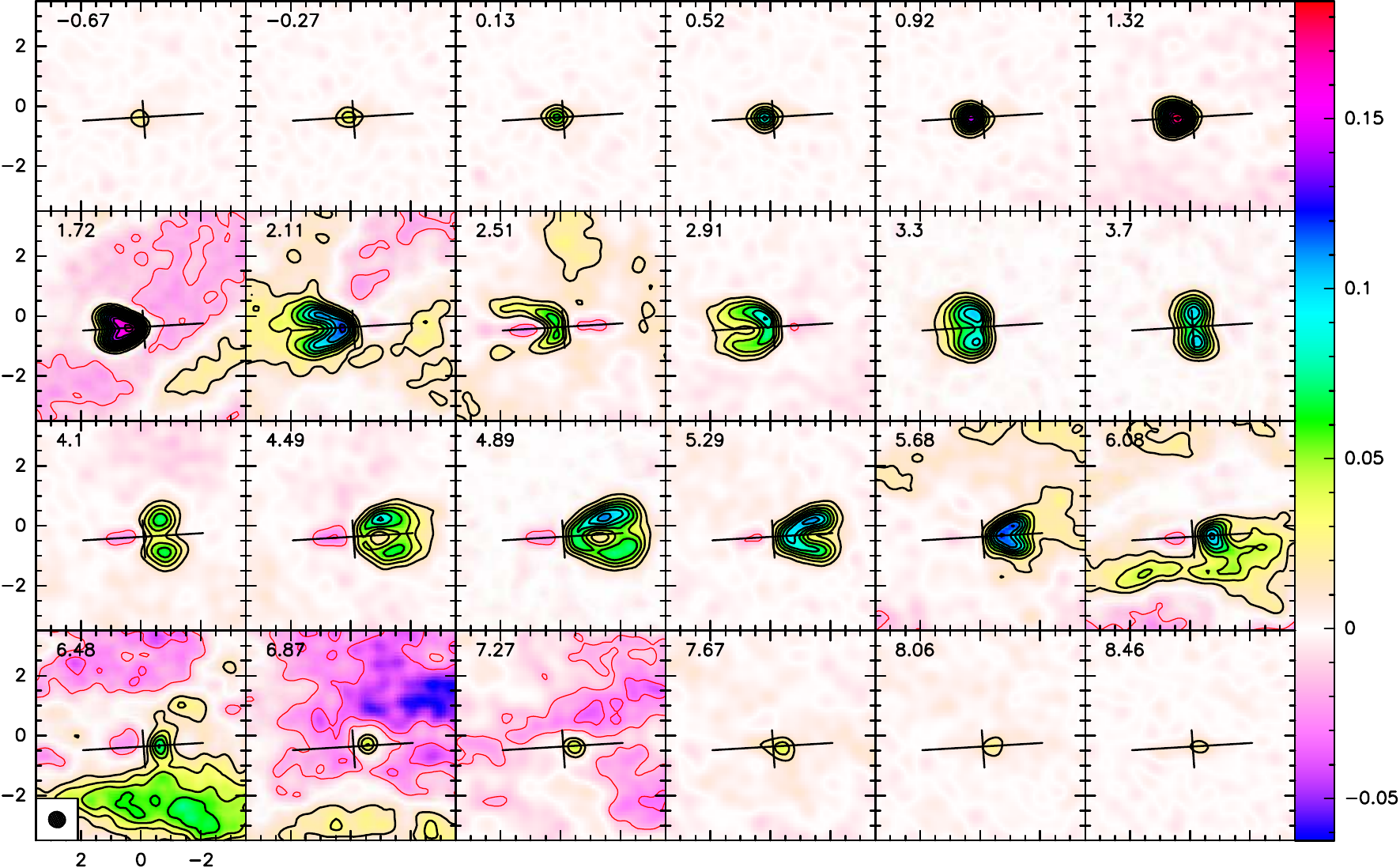}
    \caption{Channel maps of the $^{12}$CO J=2-1 line emission toward the Flying Saucer.
    Contours are in step of 15 mJy/beam (1.22 K, approx 7 $\sigma$): 
\textbf{negative contours are in red, positive contours in black, zero level contour
    omitted, and the color scale is given on the right}. The cross indicates the position and orientation of
    the dust disk. The LSR velocity (in km\,s$^{-1}$),is indicated in the upper left corner of each
    panel.}
    \label{fig:co-map}
\end{figure*}

\begin{figure*}
     \includegraphics[height=6.5cm]{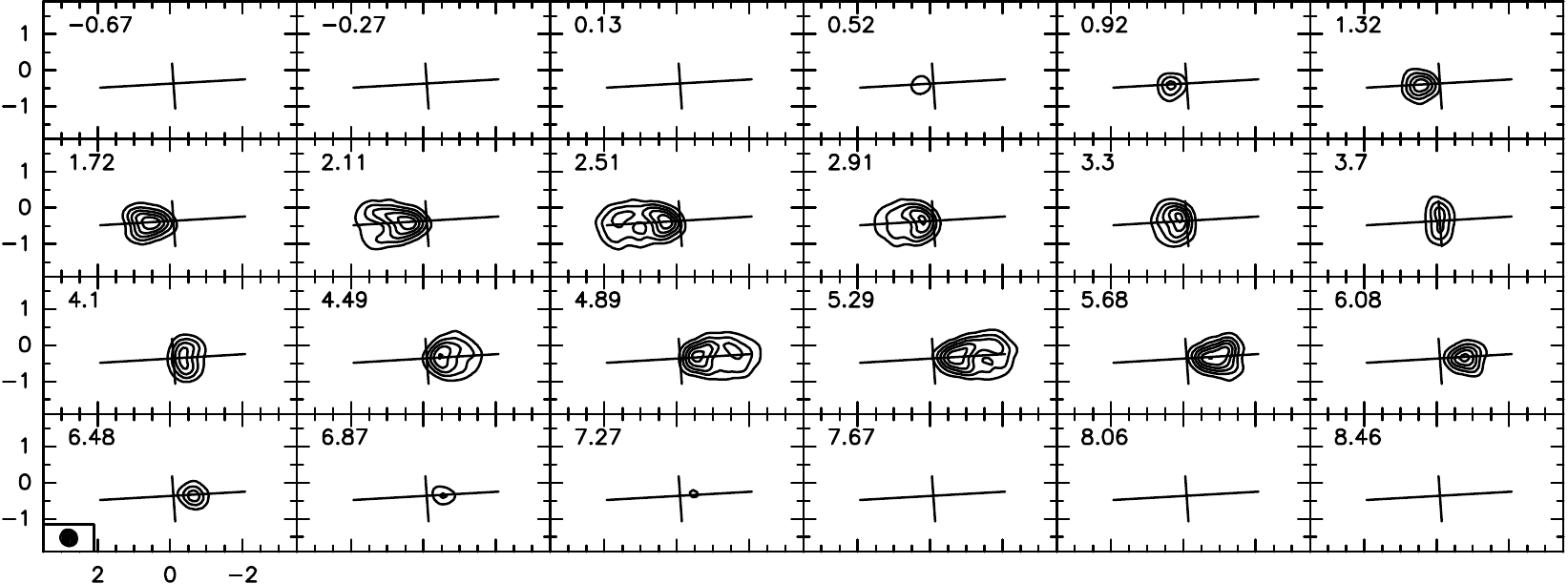}
    \caption{As Fig.\ref{fig:co-map} but for the CS J=5-4 line.
    Contours are in step of 15 mJy/beam (1.47 K, approx 9 $\sigma$). }
    \label{fig:cs-map}
\end{figure*}

Imaging observations were performed with the Atacama Large mm/submm Array (ALMA)
in a moderately compact configuration. The project 2013.1.00387.S
was observed on 17 and 18 May 2015 under excellent weather conditions.
The correlator was configured to
deliver very high spectral resolution with a channel spacing of
15 kHz (and an effective velocity resolution of 40 m\,s${-1}$).
We observed simultaneously CO J=2-1, all the most intense hyperfine components
of the CN N=2-1 transition, and the CS J=5-4 line.

Data was calibrated via the standard ALMA calibration script in the CASA software
package (Version 4.2.2). Titan was used as a flux calibrator. The calibrated data
was regridded in velocity to the LSR frame using the ``cvel'' task, and exported
through UVFITS format to the GILDAS package for imaging and data analysis.
Atmospheric phase errors were small, providing high dynamic range continuum
images and thermal noise limited spectral line data.
The total continuum flux is 35 mJy at 242 GHz (with about 7\% calibration uncertainty).
With robust weighting, the $uv$ coverage provided by the $\sim 34$ 
antennas yields a nearly circular beam size \textbf{close to  0.5$''$.
The CS images were produced at an effective spectral resolution of 0.1 km\,s$^{-1}$;
the rms noise is 3 mJy/beam, i.e. about 0.27 K given the beam size of
$0.48'' \times 0.46''$ at PA $53^\circ$. For CO, a spectral resolution of 0.08 km\,s$^{-1}$
was used, and the rms noise is 4 mJy/beam, i.e. about 0.37 K given the beam
size of $0.51'' \times 0.48''$ at PA $54^\circ$. 
}

\begin{figure*}
    \centering
     \includegraphics[width=16.0cm]{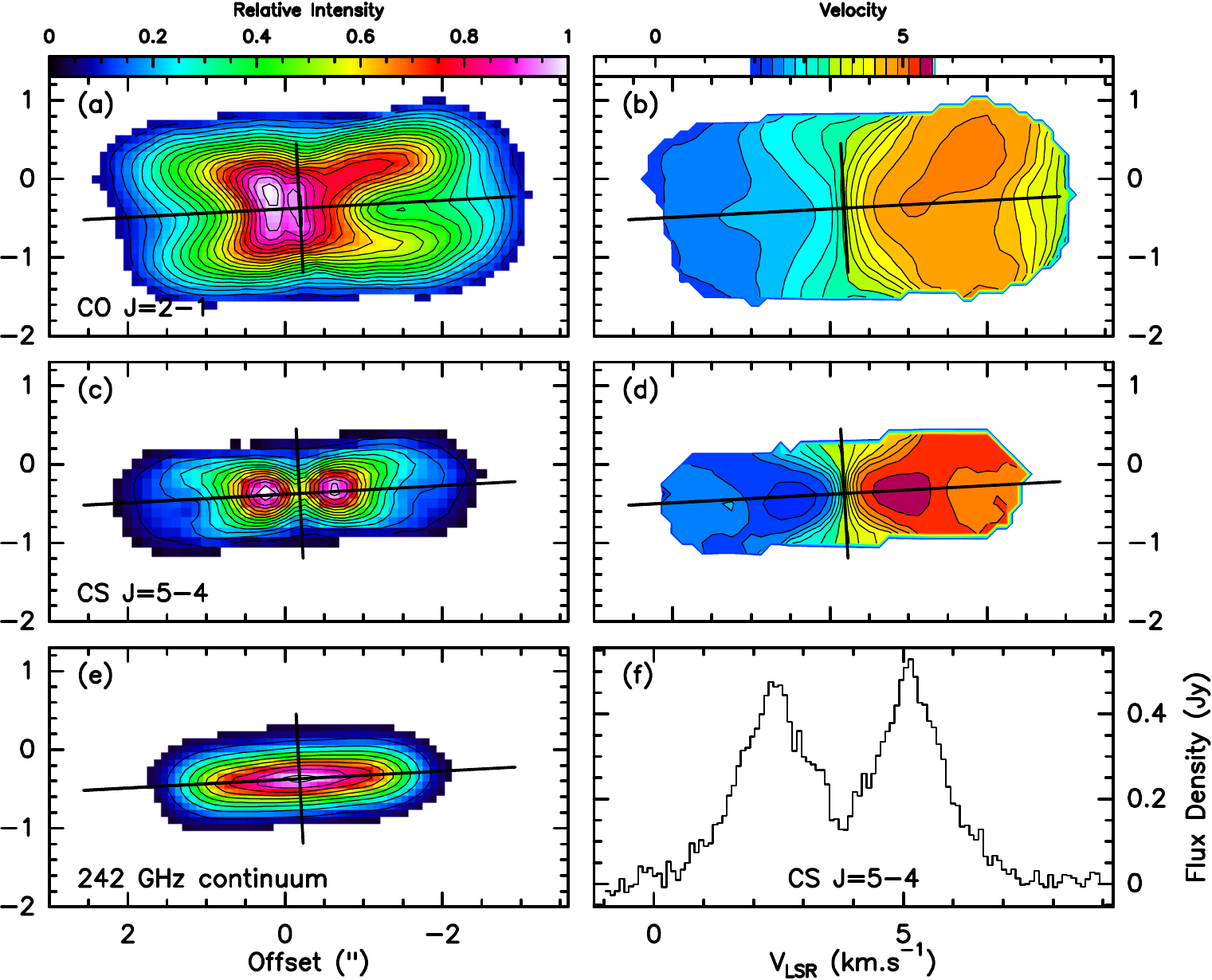}
    \caption{A summary of the observations for CO, CS and dust. 
    CO J=2-1 integrated intensity map (a) and isovelocity contours (b),
    CS J=5-4 integrated intensity map (c) and isovelocity contours (d),
    continuum image at 242 GHz (e) and CS J=5-4 integrated spectrum (f).}
    \label{fig:wholeII}
\end{figure*}

Figures \ref{fig:co-map} and \ref{fig:cs-map} present the CO and CS channel maps, respectively,
\textbf{spectrally smoothed to a $0.4\,\kms$ for clarity.} 
Figure  \ref{fig:wholeII} is a summary of the emission from CO, CS and the dust continuum.

In addition, a CO J=2-1 spectrum of the clouds along the line 
of sight was obtained with the IRAM 30-m telescope, as described in 
\citet{Guilloteau+etal_2016a}.

\section{Results}

The CO J=2-1 and continuum results were partially reported by
\citet{Guilloteau+etal_2016a}, who used them to measure the dust temperature.

\subsection{Images}

Figure \ref{fig:wholeII} clearly shows that the disk is viewed close to edge-on.
It also reveals a vertical stratification of the dust and molecules.

The modeling of the continuum image by \citet{Guilloteau+etal_2016a} leads 
for the (millimeter) dust grains to a scale height of $12.7\pm0.3$~au 
at 100 au, increasing with a $0.34\pm0.04$ exponent. For comparison, the 
modeling of the near-infrared images by \citet{Grosso+etal_2003} leads for 
the (micron) dust grains to a larger scale height of $22.5\pm1.5$\,au at 
100 au when adopting the same definition and distance, and increasing 
more rapidly with a 1.25 exponent. Therefore, there is a clear 
indication of dust settling in this disk, with large
grains preferentially close to the disk mid-plane.

The integrated intensity maps (Fig.\ref{fig:wholeII}) show that CS is significantly 
more confined towards the disk mid-plane than CO. 
As mentioned in \citet{Guilloteau+etal_2016a},
CO is contaminated by background emission from extended molecular clouds
at four different velocities, which affect the derivation of the
integrated emission and result in apparent asymmetries.
The CS molecular emission extends at least up to radius of about 300 au, 
and slightly more ($\sim 330$ au) for CO. 
On the contrary, the dust emission is confined within 200 au.
The apparent distributions may be more a result of temperature gradients,
excitation conditions and line opacities than reflecting different
abundance gradients for these molecules. The CO J=2-1 line is
much more optically thick than the CS line, and thus
more sensitive to the (warmer) less dense gas high above the disk mid-plane.
Along the mid-plane, self-absorption by colder, more distant gas, can
result in lower apparent brightness. However, at the disk edges, this
effect should be small, so the higher brightness above the disk plane
likely indicates a vertical temperature gradient, with warmer gas above the plane.

The aspect of the iso-velocity contours (Fig.\ref{fig:wholeII})
is exactly what is expected from a Keplerian flared disk seen 
edge-on. In such a configuration, at an altitude $z$ above the disk 
plane, the disk only extends inwards to 
an inner radius depending on 
$z/H$, where $H$ is the scale height, so that the maximum velocity 
reached at altitude $z$ is limited by this inner radius. Thus the mean 
velocity decreases from the mid-plane to higher altitude, resulting in 
the ``butterfly'' shape of the iso-velocity contours. The effect is 
however less pronounced for CO, as its high optical depth allows us to 
trace the emission well above the disk plane (iso-velocity contours 
would be parallel for a cylindric distribution).

\subsection{Simple determination of the disk parameters}
\label{sub:diskfit}

To constrain the basic parameters of the disk, we make a simple model 
of the CS J=5-4 emission with DiskFit \citep{Pietu+etal_2007} assuming 
power laws for the CS surface density  
($\Sigma_{CS}(r) = \Sigma_0 (r/\mathrm{100 au})^{-p}$) and 
temperature $T_{ex}(r) = T_0 (r/\mathrm{100 au})^{-q}$.
\textbf{The disk is assumed to have a sharp outer edge at $R_\mathrm{out}$.}
The vertical density profile is assumed Gaussian \citep[see Eq.1][]{Pietu+etal_2007}),
with the scale height a (free) power law of the \textbf{radius: $h(r) = H_0 (r/\mathrm{100 au})^{-h}$.
The line emission is computed assuming a (total) local line
width $dV$ independent of the radius and LTE (i.e. $T_0$ represent the
rotation temperature of the level population distribution).}

\textbf{Besides the above intrinsic parameters, the model also 
includes geometric parameters: the source
position $x_0,y_0$, the inclination $i$ and the position angle
of the rotation axis $PA$, and the source systemic velocity
$V_\mathrm{LSR}$ relative to the LSR frame.}

Results are given in Table \ref{tab:cs}. This simple model 
allows us to determine the overall disk orientation, the systemic 
velocity and the stellar mass, and gives an idea of the temperature 
required to  provide sufficient emission. \textbf{ The  apparent
scale height $H_0$ derived at 100 au would correspond to a temperature of 53 K, 
much larger than  $T_0$. The difference may indicate that CS is substantially sub-thermally
excited or, more likely, that CS emission only originates from above one hydrostatic scale height. }

\begin{table}
\caption{CS disk modeling results.}
\begin{tabular}{lcll}
\hline
Parameter & Value (at 100 au) & Unit & \\
\hline
$PA$      & $3.6 \pm 0.4$ & $^\circ$ & PA of disk rotation axis \\
$i$       & $85.4 \pm 0.5$ & $^\circ$ & Inclination \\
$V_\mathrm{LSR}$ & $3.755 \pm 0.003$ & km\,s$^{-1}$ & Systemic velocity \\
$M_*$ & $0.58 \pm 0.01$ & $\Msun$ & Star mass (a) \\
$R_\mathrm{out}$ &  $290  \pm 7$ & au & Outer radius \\
$dV$ & $0.17 \pm 0.01$ & km\,s$^{-1}$ & Local line width (b) \\
$\Sigma_0$ &  $4.3\,10^{13} \pm 0.3\,10^{13}$ & cm$^{-2}$ & CS Surface density \\
$p$        &  $2.71 \pm 0.03$  &    & Surface density exponent \\
$T_0$      &  $18.0 \pm 0.5$        & K  & CS temperature \\
$q$        &  $-0.18 \pm 0.03$    &    & temperature exponent \\
$H_0$      &  $ 25.8 \pm 0.3 $  & au & Scale height of CS (c)  \\
$h$        &  $-1.40 \pm 0.03$  &    & exponent of scale height \\
%
\hline
\end{tabular}
\\
  \label{tab:cs}
  (a) Assuming Keplerian rotation. (b) assumed constant with $r$. 
 Errors are formal errorbars from the fit. \textbf{(c) apparent scale height (see section \ref{sub:diskfit}).}
\end{table}

\subsection{PV-diagrams}
\label{sub:diagram}

\begin{figure*}
    \centering
     \includegraphics[height=12.0cm]{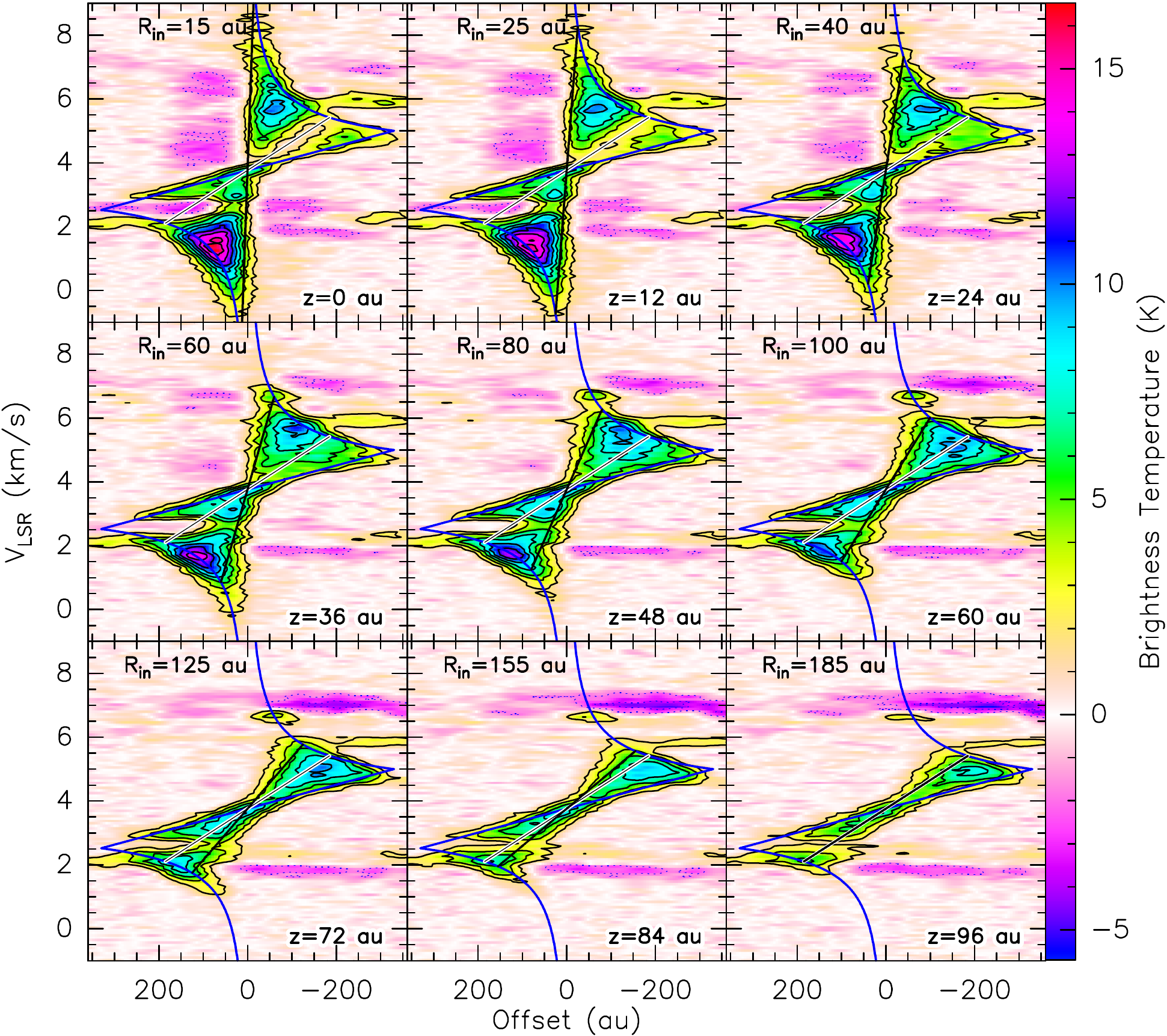}
     \includegraphics[height=12.0cm]{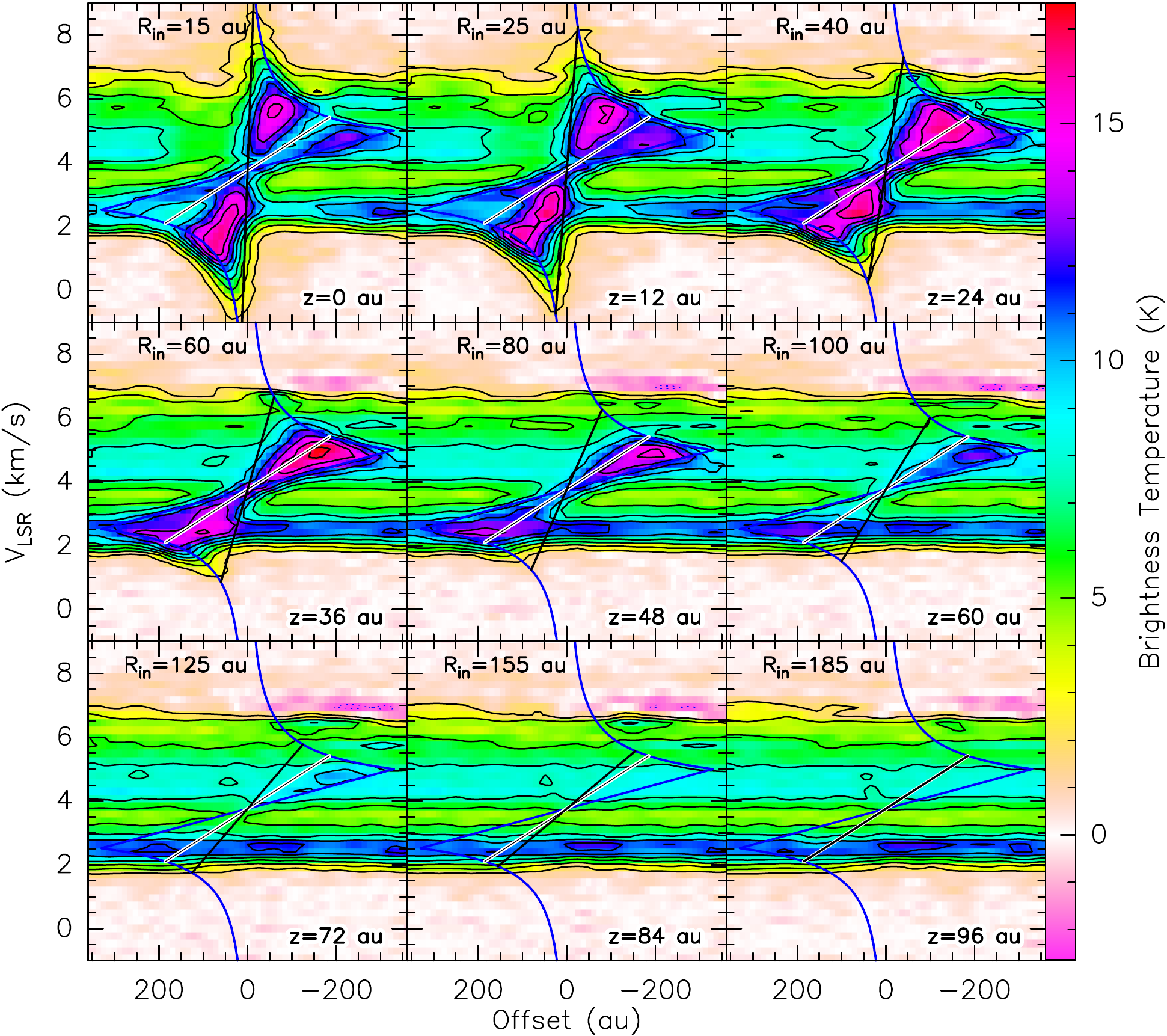}
    \caption{Position - Velocity diagram for CO J=2-1. 
    Each panel is a cut at a different height above the plane (indicated
    in the lower right corner of each panel). The spatial resolution is 
    56 au. Top: without the background added \textbf{(spectral resolution 0.08 km\,s$^{-1}$)}.
    Bottom: with the background spectrum added. On each panel, the black
    line highlights the apparent inner radius whose value is quoted in the upper
    left corner. \textbf{Here, the spectral resolution is  0.27 km\,s$^{-1}$.}    
    }
    \label{fig:co21}
\end{figure*}

\begin{figure*}
    \centering
     \includegraphics[height=12.0cm]{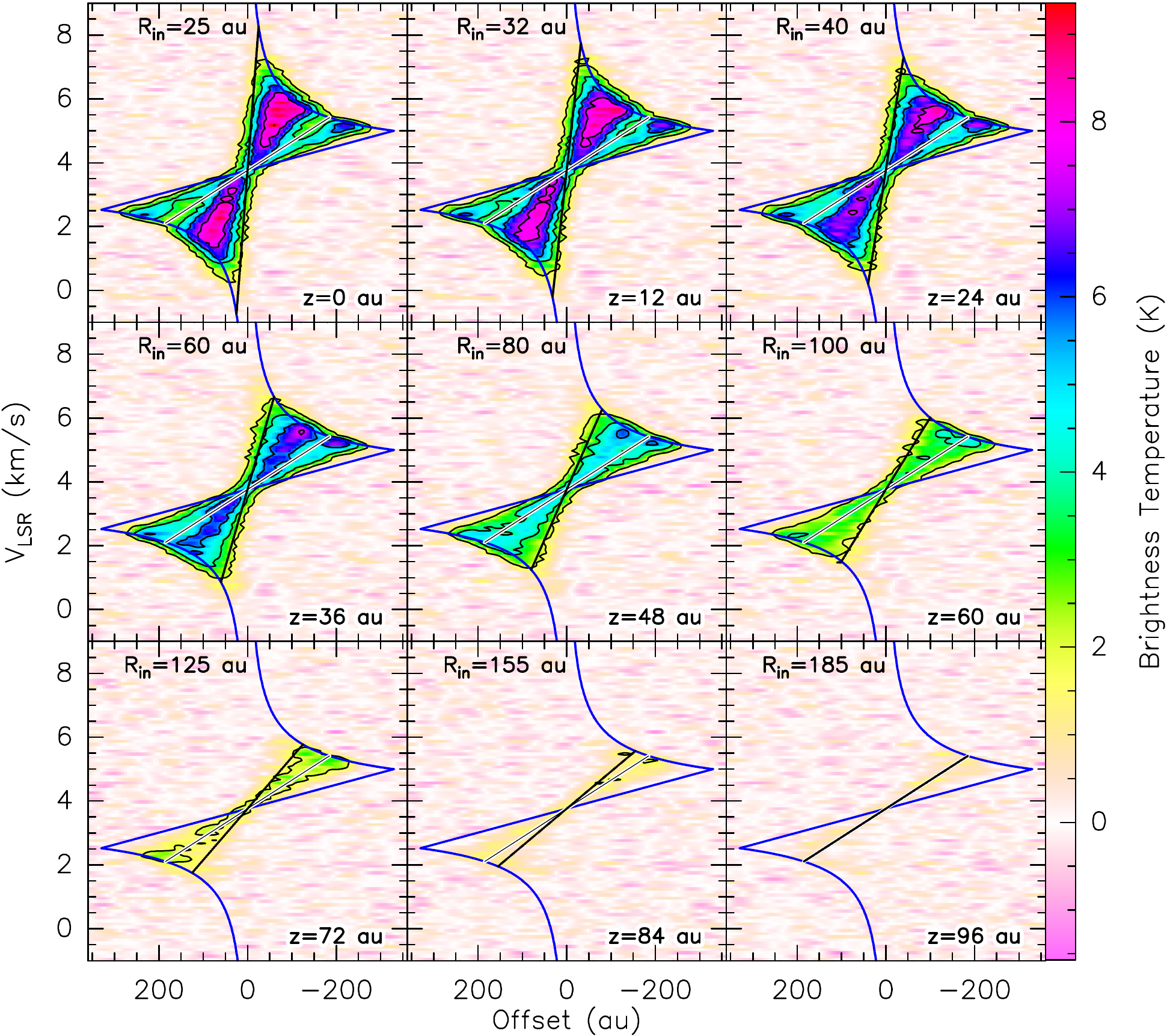}
    \caption{As in Fig.\ref{fig:co21}, but for CS J=5-4 \textbf{(spectral resolution 0.10 km\,s$^{-1}$).}
    }
    \label{fig:cs54}
\end{figure*}

A more detailed understanding of the disk properties can be derived from the 
position-velocity diagram shown in Figures \ref{fig:co21} and 
\ref{fig:cs54} where several altitudes $z$ are shown. In such diagrams, 
radial straight lines (i.e. lines with $v(x) - V_\mathrm{disk} \propto 
x$, where $x$ is the impact parameter, $v$ the velocity and $V_\mathrm{disk}$ the
disk systemic velocity) trace a constant radius $r$ (see 
Appendix \ref{app:rv} for details).  
The blue straight lines indicate the outer radius ($R_\mathrm{out} 
\simeq 330$ au). The blue curve is the Keplerian rotation curve 
$\sqrt{G M_*/r}$, with a stellar mass of $0.57 \Msun$.
The black line is the apparent inner radius $R_\mathrm{in}$, and
white-over-black line delineates the radius $r_\mathrm{dip} \simeq \rout$, 
where a dip in emission is observed at low  altitudes both in CO and in CS. 

The CO J=2-1 line provides a direct view into the thermal structure
of the disk.  Like for the continuum emission, the background provided
by the four extended molecular clouds identified in the 30-m spectrum
modulates the apparent brightness of the disk, since the ALMA array only measures
the difference in emission between the disk and the background clouds.
Because the CO J=2-1 line is essentially optically thick in 
disks,  we can simply recover a corrected CO PV diagram by adding the background
spectrum obtained with the IRAM 30m, \citep[see][their Fig.1]{Guilloteau+etal_2016a}
to the observed CO emission, at least within the disk boundaries (in position
and velocity). The result is given in Fig.\ref{fig:co21} (bottom).

In the disk plane, while CO appears to extend down to very small radii
($< 15$\,au), CS may have an inner radius around 25 au.

At a given location, the impact of the finite beamsize depends on the brightness temperature 
gradients which are different for CO and CS (because of different opacities and
excitation conditions) leading therefore to different apparent structures. 
Nevertheless, all apparent inner radii increase with height above the disk plane, as described before 
for the analysis of iso-velocity curves.
This happens because the disk is flaring due to the hydrostatic equilibrium, hence its
vertical thickness is increasing with radius.
 
The disk mid-plane is also clearly colder than the brightest background 
molecular cloud at 2.8 km\,s$^{-1}$, which has $J_\nu(T) = 11$\,K, beyond about 100 au 
radius, being almost as warm as the second brightest cloud at 4.2 km\,s$^{-1}$, 
with $J_\nu(T) = 8$\,K, at radii around $\rout$. From this simple 
consideration, we safely constrain the mean disk mid-plane temperature, 
averaged over one beam, to be 13 K near 180 au, after proper conversion 
of the brightness temperature outside of the Rayleigh-Jeans domain. It 
would rise to about 18 K at 100 au.

The PV diagrams above the disk plane indicate a warmer temperature in 
the upper layers, since no absorption is visible for $z>40$\,au or so. 
Because of our limited linear resolution (about 56 au), with such a  
vertical temperature gradient the CO PV diagram only gives an upper 
limit to the disk mid-plane temperature because the scale height (about 
10 au at 100 au) is substantially smaller than the linear resolution 
except at the disk edge.

We also note that both CS and CO shows a drop in the emission 
intensity at radius $r_\mathrm{dip} \simeq \rout$. This drop
is somewhat more difficult to identify in the CO PV diagrams
because of the background clouds. 
In CO, the emission drop seems to disappear at an height of 40 au, suggesting
it occurs only below about 30-40 au given the limited angular
resolution. Beyond a radius of 220 au, CO emission is observed again.
In CS, the deficit of emission near $\rout$ extends somewhat higher,
up to a height of 60 au.

Figure \ref{fig:co-cs-over} also reveals a north-south asymmetry, the 
north side being brighter in CS and in CO. 

\section{A new method of Analysis}

\subsection{Deriving the brightness distribution from the PV-diagram}
\label{sub:pv}

\textbf{For an homogeneous medium, the measured brightness temperature $T_b$
is given by
\begin{equation}
T_b = (1-\exp(-\tau)) (J_\nu(T) - J_\nu(T_{bg}))
\end{equation}
where 
$J_\nu$ is the Planck function multiplied by $c^2/2k\nu^2$,
\begin{equation}
J_\nu(T) = \frac{h \nu}{k} \frac{1}{\exp(h\nu/(kT))-1} . 
\end{equation} 
Since} in a PV diagram a radial line represents a constant 
radius, we can recover the disk temperature from the thermalized and 
optically thick CO J=2-1 transition by averaging \textbf{the observed
brightness} along such radial 
lines in regions where the signal is sufficiently resolved spectrally 
and spatially. \textbf{This averaging process yields the mean
radiation temperature, $J_\nu(T)-J_\nu(T_\mathrm{bg})$, from
which the temperature $T$ is derived.}
The disk being seen edge-on, cuts at various 
altitudes $z$ provide a direct visualisation of the gas temperature 
versus radius $T(r,z)$, although only at the angular resolution of the 
observations. \textbf{The 2-D image resulting from the application
of this averaging process is called hereafter the \textit{tomographically
reconstructed distribution} or TRD}.

\textbf{While the CO J=2-1 TRD is just the temperature distribution},
for optically thinner \textbf{or non thermalized} lines, the interpretation is 
more complex because the \textbf{TRD} is a function of both the 
temperature and local density. Nevertheless, it also provides a direct 
measurement of the altitude of the molecular layer. 

\begin{figure*}
    \centering
     \includegraphics[height=10.0cm]{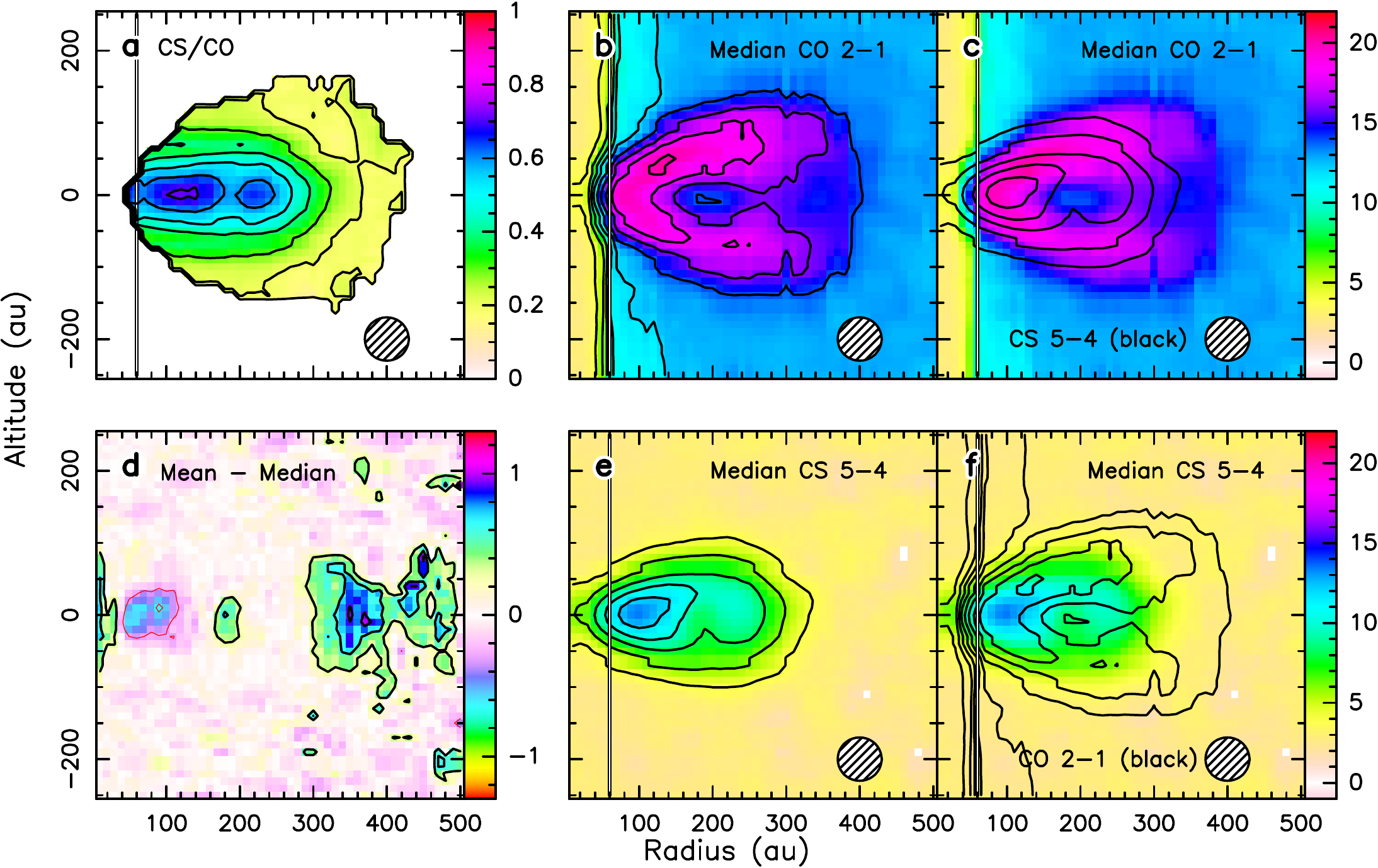}
    \caption{Derivations of the gas temperature from the PV-diagram. 
    (a) ratio of the median \textbf{TRD} 
    derived from CS over the median \textbf{TRD} derived from CO 
    (contours are 0.1 to 1 by 0.1) (b) CO median \textbf{TRD}, with  
    contours from 4 to 22 K by spacing of 2 K. (c) CO median 
    \textbf{TRD} with  contours of the CS median \textbf{TRD} overlaid.
    (d) difference between the CS mean temperature and the CS median 
    \textbf{TRD} (contours are from -0.6 to 0.6 K by spacing of 0.3 K).
    (e) median CS \textbf{TRD}  with contours from 4 to 12 K by 
    spacing of 2 K. (f) CS median \textbf{TRD} with contours of the 
    CO median \textbf{TRD}. The spatial resolution is indicated. 
    The vertical line delineates the area suffering from beam dilution.}
    \label{fig:co-cs-over}
\end{figure*}

We use the PV-diagrams of CO and CS transitions to derive \textbf{their
respective TRDs}
and show them in 
Fig.\,\ref{fig:co-cs-over}. The derivations were performed onto data 
cubes without continuum subtraction because the subtraction, which was 
needed to compute the iso-velocity contours, may lead to substantial 
problems near the peak of the continuum. For CO, which is 
optically thick, the TRD$(r,z)$ map is obtained using the PV-diagram with 
the CO emission from the background clouds added, since that emission 
is fully resolved out by the ALMA observations. This explains why the 
derived CO temperature brightness is of the order of 15~K  around the 
disk, in agreement with the values derived from the CO spectrum taken 
with the IRAM 30-m radiotelescope \citep{Guilloteau+etal_2016a}. 

In each case, we calculated a map of the mean, median and  
maximum brightness along each radius. For CO, the mean and maximum 
brightness are contaminated by the background clouds, and the median is 
expected to be a better estimator. Furthermore, for CS, we found that 
the mean and the median always give results which differ by less than 1 
K inside the disk at $r < 350$~au (see Fig.\ref{fig:co-cs-over}, panel (d) 
which shows the difference between the mean and the median 
for CS 5-4). This indicates that the derivations are robust and do not 
suffer from significant biases. The observed \textbf{TRDs}  
clearly confirm the location of the molecular layer above the mid-plane 
but the CS emission  peaks at a lower temperature and is located below 
the CO emission. The ratio of the CS \textbf{TRD} over the 
CO \textbf{TRD} confirms these behaviors  
(see panel (a) of Fig.\ref{fig:co-cs-over}).

The Keplerian shear implies that the spatial averaging of the derived 
brightness is not the same everywhere inside the disk. Indeed, at a 
radius $r$ corresponding to a velocity $v(r)$, the smearing due to the 
Keplerian shear is given by  $dr = 2r dv/v(r)$ where $dv$ is the local 
linewidth, which is due to a combination of thermal and turbulent 
broadening. This limits
the radial resolution which can be obtained in the disk outer
parts. For instance, for CO, at 20 au, assuming a temperature of 30 K and 
a turbulent boadening ($0.05 \kms$) similar to that observed in TW Hydrae 
disk by \citet{Teague+etal_2016}, the $dr$ 
would be of the order of $\sim 2$~au, a value which would
not affect studies at spatial resolution down to $0.1''$ (12 au) or so.
On the contrary, at radius 200 au, the $dr$ would be of the order 
of 35 au for the same line width, dropping to 20 au assuming the 
same turbulence but  a (mid-plane) temperature of 7 K. This limits 
the gain obtained with high angular resolution only at large radii.
It also explains why the apparent extent of the \textbf{tomographically reconstructed
distribution} in Fig.\ref{fig:co-cs-over} exceeds the
disk outer radius more than expected from the angular resolution
only.
 
Nevertheless, this smearing is 
purely radial and the vertical structure is not affected: 
the smearing is the same above and below the mid-plane at a 
given radius (assuming there is no vertical temperature gradient).

This occurs because the disk is edge-on but it is no longer true
in less inclined disks. In such disks, the smearing resulting from
the local line width will limit the effective resolution radially
and vertically.

Finally, we note that this direct method of analysis, specific to 
edge-on disks, is complementary to the classical channel maps studies by 
providing a more synthetic but direct view of the vertical disk 
structure.      

%
 

\subsection{CO Modeling using DiskFit}
\label{sect:mod}

To go beyond the resolution-limited information provided in
Fig.\ref{fig:co-cs-over}, we study here the impact of 
several key parameters of the disk by performing grids of models to 
better constrain the disk geometry and structure.  We use the 
ray-tracing model DiskFit \citep[see Section \ref{sub:diskfit},][]{Pietu+etal_2007}. 
For simplicity, we 
assume LTE conditions, which is appropriate for CO.
\textbf{The temperature structure is here more complex than
the simple vertically isothermal model of Section \ref{sub:diskfit}.}
The atmosphere temperature is given by
\begin{equation}
T_{atm}(r) =  T_{atm}^0 \left(\frac{r}{r_0}\right)^{-q_{atm}}
\end{equation}
and the mid-plane temperature is given by 
\begin{equation}
T_{mid}(r) =  \min \left(T_{atm}(r),T_{mid}^0 \left(\frac{r}{r_0}\right)^{-q_{mid}}\right)
\end{equation}
In between for an altitude of $z < z_q$ , the temperature is defined by 
\begin{equation}
T(r) = \left(T_{atm}(r)-T_{mid}(r)\right) \left(\cos\left(\frac{\pi z}{2 z_q H(r)}\right)\right)^{2\delta}  + T_{mid}(r)  
\end{equation}
where $H(r)$ is the hydrostatic scale height (defined by $T_{mid}(r)$).
Provided $q_{atm} > q_{mid}$, there is a radius $R_q$ beyond
which the temperature becomes vertically isothermal.
Note that for $q_{mid} = 0$, this definition is identical to that used by 
\citet{Dartois+etal_2003}. 

The models of the molecular emission were performed together with the 
continuum emission not subtracted. The continuum model uses, in particular 
for the density and temperature, 
the parameters defined in Table 1 of \citet{Guilloteau+etal_2016a}. 
We take into account the 
spatial resolution by convolving all models by a $0.5''$ circular beam.
In all models, the outer radius $r_{out}$ is taken at 330 au, in 
agreement the value derived from the PV-diagram.

We find that a small departure from edge-on inclination by 2-3 degrees 
is sufficient  to explain the small North-South brightness asymmetry
visible in Fig.\ref{fig:co-cs-over}. The  most probable value is 
$i = 87^\circ$ (see Sect.\ref{sec:disk-structure}), a value
used for all further models.

In a first series of models, we assume a CO vertical distribution which 
follows the H$_2$ density distribution, i.e. assuming a constant 
abundance. For the H$_2$ density distribution, we assume either 
power laws or exponentially tapered distribution  
following the prescription given in \citet{Guilloteau+etal_2011}.
We explored CO surface densities ranging from 10$^{16}$  to 10$^{19}$ 
cm$^{-2}$ at 100 au.  To 
account for the observed brightness at high altitudes above the disk 
plane,  we find that the CO surface density at 100 au must be at least 
of the order of 5~$10^{17}$ cm$^{-2}$, with a power law  index of the 
order of $p=1.2$ for the radial distribution.  We also explored the 
impact of the temperature distribution $z_q$, $\delta$, $T_{atm}^0$, 
$q_{atm}$,  $T_{mid}^0$, $q_{mid}$ and $R_q$. Best runs are obtained for 
mid-plane temperatures $T_{mid} \approx 10$ K at 100 au,  $q_{mid} = 
0.4$ leading to about 6 K at the outer disk radius and 17 K at 26 au (CO 
snowline location),  $T_{atm}^0 = $50~K, $q_{atm}=0-0.2$, $\delta = 2$ 
and $z_q = 1.3 - 2$.   

 
However none of these models, which assume CO molecules 
are present everywhere in the disk, properly reproduce the CO depression 
observed around the mid-plane (see panel (b) of 
Fig.\ref{fig:co-cs-over}). This remains true even if the mid-plane 
temperature is set to values 5-7\,K, matching
the temperature of large grains measured by \citet{Guilloteau+etal_2016a}.

We attempted to perform more realistic models by assuming 
complete molecular depletion (abundance $\xpco  = 0$) in the disk mid-plane. In 
this model, the zone in which CO is present is delimited upwards by a 
depletion column density $\Sigma_{dep}$ and downwards by the CO 
condensation temperature. CO molecules are present (with a constant 
abundance $\xuco =10^{-4}$) when the H$_2$ column density from the current $(r,z)$ point 
upwards (i.e. towards ($r,\infty$)) exceeds a given threshold 
$\Sigma_{dep}$, to reflect the possible impact of photo-desorption of 
molecules, or when the temperature $T(r,z)$ is above 17\,K.
Such a model takes into account the possible presence of CO in the 
inner disk mid-plane inside  the CO snowline radius,  and also at large 
radii because as soon as the surface density becomes low enough,
CO emission can again be located onto the mid-plane.       
In this model, the CO surface density radial profile 
$\Sigma(r) \times \xuco $ (where $\Sigma(r)$ is the H$_2$ surface
density profile) is constrained, because of the need to provide sufficient 
opacity for the CO J=2-1 at high altitudes above the disk 
mid-plane to reproduce the observed brightness. The derived H$_2$ densities are then strictly 
inversely proportional to the assumed $\xuco$.

Table \ref{tab:model} gives the
parameters of the best model we found with this approach.
\textbf{Unfortunately the \textbf{current} angular resolution of the data 
limits the analysis. Because of this limited angular resolution,
parameters $T_{atm}^0,\delta,z_q$ are strongly coupled.
$z_q$ also depends implicitely on  $T_{mid}$, because
it is the number of hydrostatic scale heights at which the atmospheric
temperature is reached. In practice, Table \ref{tab:model} only
confirms a low mid-plane temperature ($10$ K at 100 au), 
and temperatures at least a factor 2 larger than this in the CO rich region,
consistent with the (spatially averaged) values derived in Section
\ref{sub:diagram}.
Only higher spatial resolution data would allow us to break the
degeneracy and accurately determine the vertical temperature gradient
(see Section \ref{sub:spatial}).
}

Furthermore, even this model only qualitatively reproduces the 
brightness distribution of the CO emission around the mid-plane. In 
particular, the  shape of the depletion zone is difficult
to evaluate from the current data. We also fail to reproduce the rise of the CO brightness 
after 250 au, most likely because our model does not include an increase of 
the temperature in the outer part. 

Most chemical models \citep[e.g.][]{Reboussin+etal_2015} predict that there is
some CO at low abundance ($\xpco \sim 10^{-8} - 10^{-6}$) in the 
mid-plane, depending on the grain sizes and  the local dust to gas ratio,
contrary to our simple assumption of $\xpco =0$. Such low abundances would
not impact our determination of the CO surface density, which only relies
on the need to have sufficient optical thickness in the upper layers.
However, CO could start being sufficiently optically thick around
the mid-plane, diminishing the contrast between the mid-plane and the
molecular layer. We estimated through modelling that this happens  for $\xpco  \geq 3\,10^{-8}$.
In such cases, the brightness distribution becomes similar to that of
the undepleted case. Observations of a
less abundant isotopologue would be a better probe of the mid-plane depletion.

We thus conclude that the current data are insufficient to 
disentangle between molecular depletion and a very cold mid-plane, 
but indicate a rise in mid-plane temperature beyond 200 au.

\begin{figure}
    \centering
     \includegraphics[height=8.0cm]{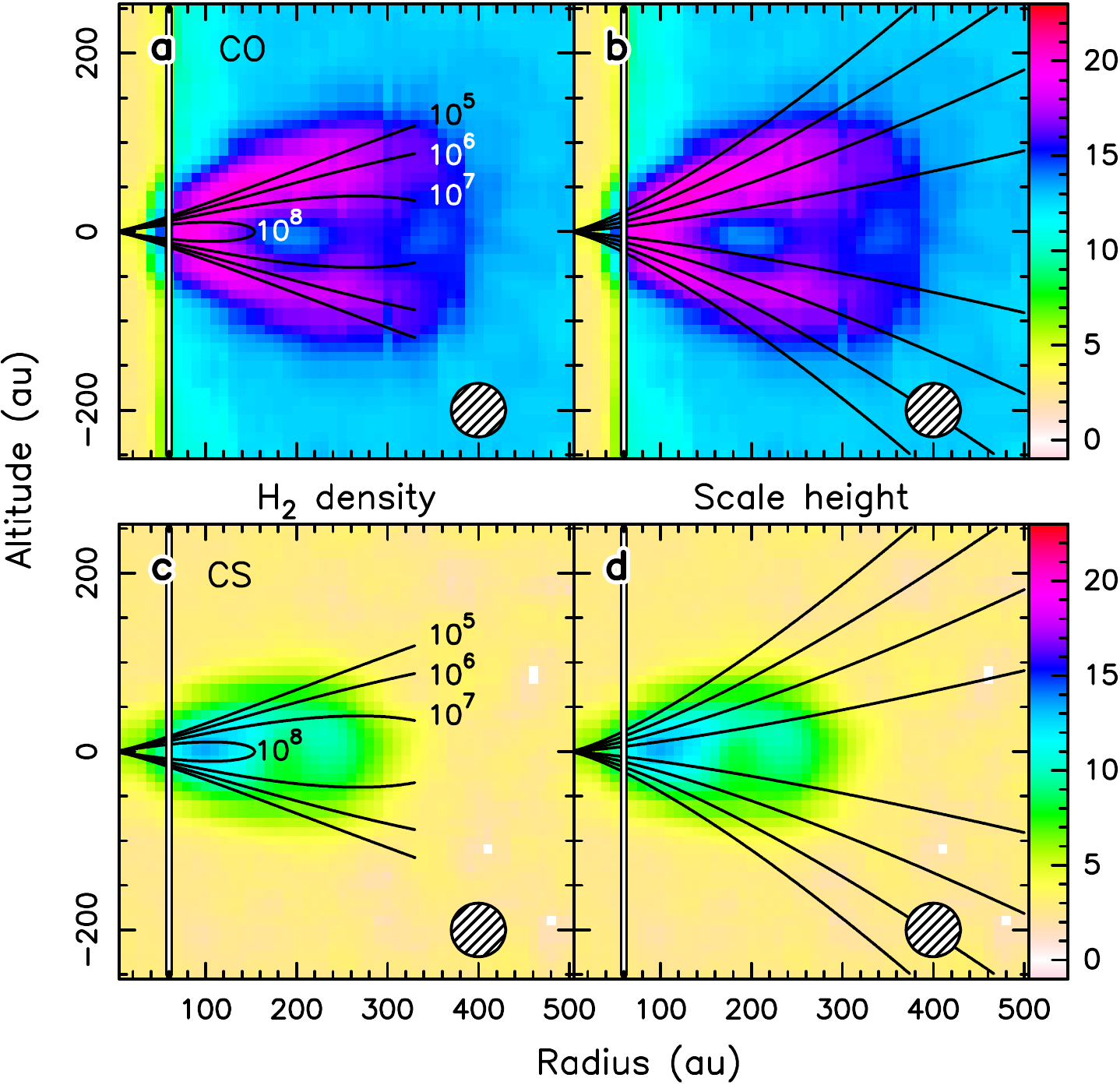}
    \caption{Superimposition of the CO and CS median \textbf{TRD} 
    to a standard disk model. (a): CO \textbf{TRD (colour)}  and 
    the H$_2$ volume  density superimposed in  \textbf{labelled} contours from $10^{5}$ up to $10^{8}$ 
    cm$^{-3}$. (b): CO \textbf{TRD} with the scale height 
    superimposed in black contours from 1 to 4 scale heights. (c) and (d)
    as (a) and (b) but for CS. The model corresponds to the parameters given in the 
    Table \ref{tab:model}.}
    \label{fig:structure}
\end{figure}

\begin{figure*}
    \centering
    \includegraphics[width=18.0cm]{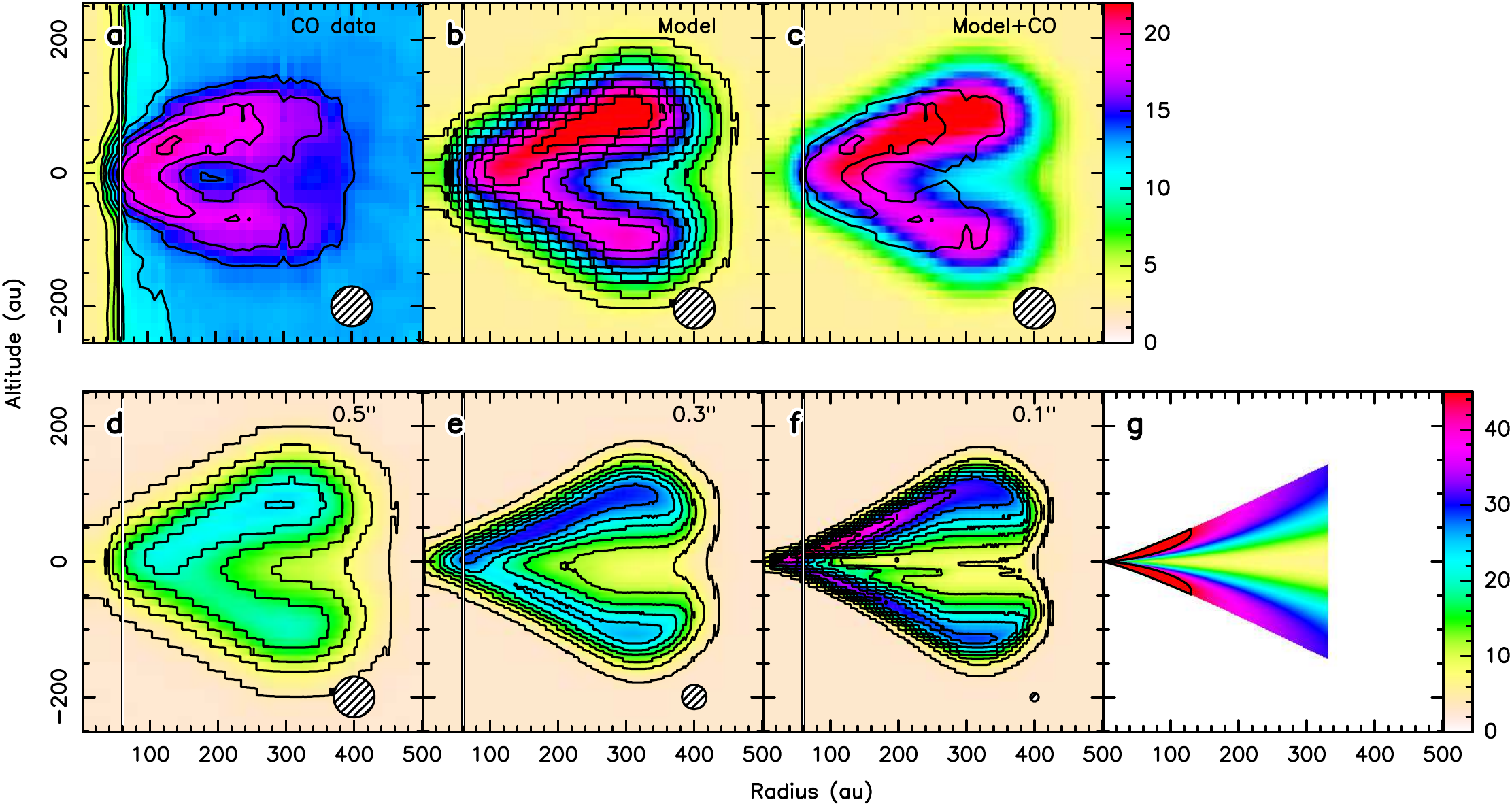}
   \caption{
   \textbf{TRD  for CO J=2-1 (in Kelvin). 
   (a) observed CO TRD. (b) best model. (c)
   CO TRD in false color with best 
   model in black contours. (d) model at 
   $0.5''$ or 60 au resolution, (e) at $0.3''$ or 36 au 
   resolution and (f)  at $0.1''$ (12 au)  
   resolution, presented with the same color scale.
   Contours are in 4 K steps.
   (g) Initial temperature distribution;
   the black contour is at 45 K.}
   }
  \label{fig:resol}
\end{figure*}

In Fig.\,\ref{fig:structure}, we overlay the brightness 
temperature derived from the observations to the structure
of the model given in Table\ref{tab:model}. The brightness
temperature from the model is compared to the observations
and also shown in Fig.\,\ref{fig:resol}  at three different 
angular resolutions of $0.5'', 0.3''$ and $0.1''$. 

\begin{table}
\caption{Model derived from the CO observations}
\begin{tabular}{llll}
\hline
Parameter & Value & Unit & Parameters at 100 au \\
\hline
$T_{atm}^0 $      & 50  & K  & Atmosphere temperature \\
$q_{atm} $      & 0.4  &   &    \\
$z_q $      & 3 &  &  \\
$\delta $      & 2 &  &  \\
$T_{mid}^0 $      & 10 & K  & Mid-plane temperature \\
$q_{mid} $      & 0.4  &  &  \\
$\Sigma_0$ &  $10^{23}$   & cm$^{-2}$  & H$_2$ Surface density \\
$p$        &  $0.5 $  &    & Surface density parameter \\
$R_{C}$        &  $50 $  & au   & Radius for exponential decay \\

$H_0$      &  11.3  & au & Scale height  \\
$h$        &  -1.3  &    & exponent of scale height \\
$R_\mathrm{out}$ &  $330 $ & au & Outer radius \\
$i$        & $87$ & $^\circ$ & Inclination \\
$\Sigma_{dep}$ &  $10^{22}$    &  cm$^{-2}$  & Surface density for depletion \\
$T_{dep}$ &  17   & K  & Depletion temperature  \\
$X_u($CO$)$ & 10$^{-4}$ & & CO abundance in upper layers \\
$X_p($CO$)$ & 0 &  & as above, in mid-plane \\
$M_\mathrm{disk}$ & $0.7\,10^{-3}$ & $\Msun$ & Disk mass \\
\hline
\end{tabular}
  \label{tab:model}
  \tablefoot{$\Sigma(r)$ corresponds to Eq.\,5 in \citet{Guilloteau+etal_2011},
 $H(r) = H_0 (r/100\,\mathrm{au})^{-h}$, and $T(r)$ is defined in Eq. 1 to 3, this paper.
 The disk is in Keplerian rotation around a 0.57 $\Msun$. Note that only the 
 product of $\Sigma_0 \times X_u($CO$)$ is constrained by the observations: 
 changing  $X_u($CO$)$ will change $\Sigma_0$ and $M_\mathrm{disk}$ by the inverse amount.}
\end{table}
 
\section{Discussion }

\subsection{Overall disk structure}
\label{sec:disk-structure}

The analysis of the CO and CS brightness temperature patterns shows 
that there is no significant departure from a simple disk geometry at 
the linear resolution of 60 au, with the exception of the North-South 
brightness difference.  This may be due to an intrinsic assymetry. 
However, the same effect can also be produced if the disk is slightly 
inclined by a few degrees from edge-on. Due to the flaring and radial 
temperature gradient, the optically thick emission of the far side 
always originates from slightly warmer gas 
\citet{Guilloteau+Dutrey_1998}.  Indeed, we find that an inclination 
angle of 87$^\circ$ is enough to account for the North-South dissymmetry, 
with the Southern part of the disk being closer to us. This inclination 
(and orientation) is in very good agreement with that of $86 \pm 1^\circ$ 
derived by \citet{Grosso+etal_2003} from near InfraRed (NIR) 
observations. The brightness distributions of the NIR images also show 
that the southern part of the disk is closer to us. Finally, there is 
no apparent sign of warp beyond a radius of about 50 au.


Figure \ref{fig:structure} shows the CO emission is 
above one scale height, while the CS emission appears slightly below and 
less extended vertically. This is not surprising. CS J=5-4 is excited only 
at the high densities near the mid-plane while CO J=2-1 is easy to 
thermalize at density as low as a few 10$^{3}$ cm$^{-3}$ (see 
panels a-c of Fig.\ref{fig:structure} which trace the density 
distribution). As a consequence, the convolution by the $\sim 56$\,au 
beam leads to different positions of the peak of the emission layer. 
The CO layer is vertically resolved and extended, while the CS emission 
is vertically unresolved  and peaks just below one scale height. These 
vertical locations for the CO and CS layers are consistent with 
predictions by chemical models \citep[see][Fig.\,8 where CS peaks between 1 and 2.5 scale 
heights]{Dutrey+etal_2011}.  A CO layer above the mid-plane was already 
observed in the disk of HD\,163296 
\citep{Gregorio-Monsalvo+etal_2013,Rosenfeld+etal_2013}. 

We also estimate for the first time the amplitude of the vertical 
temperature gradient between the cold mid-plane, the molecular layer 
and CO atmosphere at radii between  50 and 300 au. The gas vertical 
temperature gradient which is derived at 100 au is in agreement with 
that predicted by thermo-chemical models  
\citep[e.g.][]{Cleeves+etal_2016} with a mid-plane at about 8-10 K and a 
temperature of 25 K reached at two scale-heights. The main limitation 
here is the linear resolution of 60 au. Nevertheless, the observed pattern of the 
brightness distribution suggests the existence of warmer gas inside a 
radius of about 50-80 au. Any inner hole of radius $<15$\,au cannot be 
seen in these data  due to our sensitivity limit.
 
\subsection{Radial profile of CO near the mid-plane} 

\paragraph{Rise of the CO brightness a large radius:}

While in the upper layers, the CO brightness decreases smoothly with 
radius, in the mid-plane, we observe a rise in CO brightness beyond a 
radius of about 200 au. The transition radius coincides with the 
mm-emitting dust disk outer boundary, $\sim 187$ au in the simple model 
from \citet{Guilloteau+etal_2016a}, while in scattered light, the disk 
is nearly as extended as the CO emission \citep{Grosso+etal_2003}. This 
suggests that a change of grain size distribution may also play a role 
here. With the dust composed of mostly micron-sized grains 
better coupled with the gas \citep[see][]{Pontoppidan+etal_2007}, more 
stellar light may be intercepted, resulting in a more efficient heating 
of the gas, as suggested by recent chemical models 
\citep{Cleeves+2016}. Moreover, at the expected densities in the disk 
mid-plane, the dust and gas temperatures should be strongly coupled, 
and a dust temperature rise is also naturally expected when the disk 
becomes optically thin to the incident radiation and for the 
re-emission, as shown e.g. by \citet{Dalessio+etal_1999}.

Also, as the $\rho$ Oph region is bathed in a higher-than-average UV field due 
the presence of several B stars, this effect may be reinforced by 
additional heating due to a stronger ambient UV field. However, the 
Flying Saucer is located on the Eastern side of the dark cloud, 
whose dense clouds absorb the UV from the B stars, located 
mainly on the Western side. Fig. 1a and 4a of \citet{Lim+2015} show 
that there is a clear dip in the FUV emission due to the dark cloud 
silhouette. Therefore at the location of the Flying Saucer (l=353.3, 
b=16.5) the ambient FUV field cannot be very large.

\paragraph{Apparent gap at radius $\rout$:}

The presence of an apparent gap at $\sim$ 185 au both in CO and CS is 
puzzling. Its existence in CO however indicates that a change in temperature 
and/or beam dilution is the primary cause for this deficit, as CO is
easily thermalized and optically thick. This  suggests 
that the disk mid-plane warms up beyond this radius, before the emission 
fades again near the disk edge (of the order of  $\sim$ 330 au), where CO 
becomes optically thin and CS unexcited.

An alternative explanation is that it is the result of real gap 
in the molecular distribution, smoothed out by our limited angular 
resolution. The optically thick CO line traces material both at low and high 
altitude (up to 3-4 scale heights above the mid-plane) while CS J=5-4, 
a high density tracer, is more optically thin and only observed in high 
density regions (at typically one scale height). Contrary to a face-on or inclined disk, the gap 
can be seen in $^{12}$CO because the disk is edge-on. 

The brightness minimum at $\rout$ is only seen at low 
altitudes, typically between the mid-plane and an altitude of 40 au in CO,
and slightly higher in CS, so 
the putative gap cannot extend up in the disk. This morphology
remains consistent with expectations for gaps created by planets,
provided the Hills radius is smaller than the disk scale height.

\subsection{Deriving the gas density from CS excitation conditions}

In CS J=5-4, we observe a similar North-South asymmetry than in CO J=2-1.
This indicates that the emission has a substantial optical thickness
along the line of sight. 

The brightness ratio of CS over CO is within the range  0.5-0.7 until a 
radius of about 250 au and drops quickly beyond. As indicated 
by the iso-density contours shown in Figure \ref{fig:structure}, this 
is consistent with the decreasing H$_2$ density with radius, because 
the CS J=5-4 transition is thermalized at a few $10^6$ cm$^{-3}$ 
\citep{Denis-Alpizar+etal_2017}.

On the contrary, the observed ratio of 0.5-0.7 in the inner disk 
mid-plane cannot be explained by excitation conditions. A simple LVG 
calculation using the CO freeze out temperature $T_k = 17$\,K 
(which is in agreement with the CS temperature derived from the simple 
analysis),  a surface density of $4\,10^{13}$\,cm$^{-2}$ and a local 
line width of 0.3 km\,s$^{-1}$ (obtained from a simple analysis with 
DiskFit using a power law surface density distribution) yields $T_{ex} 
= 12$\,K for a density of $10^6$ cm$^{-3}$. This \textbf{density} is much lower than 
expected in the disk given the dust emission observed at mm 
wavelengths, and than the values derived from our CO modelling. 

The ratio is better explained as resulting from different beam 
dilutions in CO and CS. At 200 au, while the CO emission is spatially 
resolved, the region emitting in CS J=5-4 must fill only 50\% of  the 
synthesized beam, i.e. must have a thickness of only $\sim 30$\,au. The 
emission peak being located $\sim 45$ au above the plane at this 
radius, the CS emitting layer must be confined between about 30 and 60 
au there.  
Furthermore, the density at $r=200$\,au and $z=60$\,au is 
about $10^6$ cm$^{-3}$ in our fiducial disk model, so that the upper 
layers are no longer dense enough to excite the J=5-4 line. Hence the 
observed CS/CO brightness ratio is roughly consistent with a molecular 
layer extending above  one scale height, with sub-thermal excitation of 
the CS J=5-4 transition truncating the CS brightness distribution 
upwards.

In this interpretation, the CS layer is expected to be thinner for 
higher J transitions. An angular resolution around $0.2''$ would be 
needed to resolve the CS layer.

At smaller radii, the CO and CS layer become both unresolved 
vertically, but the ratio of their respective thickness is not expected 
to change significantly, leading to the nearly constant brightness 
ratio.


\subsection{Limits: angular resolution, local line width
and inclinations.}
\label{sub:spatial}

Figure \ref{fig:resol} shows the TRD of the 
model obtained with DiskFit assuming CO depletion around the mid-plane 
at 3 different angular resolutions, $0.5''$, $0.3''$ and $0.1''$. The 
impact of the resolution on this apparent brightness distribution is 
striking. 
\textbf{For comparison, the intrinsic temperature $T(r,z)$ 
distribution is shown in panel (g) for all points where the H$_2$
density exceeds $10^4$\,cm$^{-3}$.
Comparing panels (f) and (g) shows the impact of the small
Keplerian shear compared to the local line width
at the disk edge (which spreads the TRD beyond the outer disk radius, 
see Section \ref{sub:pv}), 
and of the small deviation from a pure edge-on disk (which result
in a top/bottom asymmetry).}
For this specific disk structure, the peak brightness is 
lowered by a factor 2 when degrading the angular resolution from 
$0.1''$ to $0.5''$. The inner CO disk, the radius of the CO snowline 
and the whole gas distribution can be resolved at $0.1''$, but not at 
$0.5''$. The stratification of the molecular layer can only be studied 
at $0.1''$ resolution, \textbf{down to about 30 au, where the scale
height becomes too small compared to the linear resolution for a direct measurement}.
We also observe a displacement of the peak 
brightness towards larger radii at $0.5''$ resolution compared to its 
(true) location at $0.1''$ resolution. This is due to the fact that the 
vertical extent of the CO emission at large radii is larger due to the 
flaring. Finally, at $0.5''$, we find that it is not possible to 
determine the shape of the (partly) depleted zone around the mid-plane 
beyond the CO snowline radius. Even a resolution of $0.3''$ would allow 
the measurement of the depletion factor (and estimate of the CO/dust 
ratio) around the mid-plane while an angular resolution of $0.1''$ 
would in addition provide the determination of the shape of this area.

\textbf{Besides angular resolution and local line width,
inclination is another limitation of the method. To first order,
the disk should be edge-on to within about $h(r)/r$ for the TRD
to be directly useable, i.e. in the range $80-90^\circ$ given
the typical $h(r)/r$ of disks at 100-300 au. 
However, even for somewhat lower inclination, the TRD can give 
insight onto the location of the molecular layer.
We illustrate this in Appendix \ref{app:incli}.}

Such simulations clearly demonstrate how powerful ALMA can be to 
characterize the structure of an edge-on protoplanetary disk, provided 
its distance is reasonable. 


\section{Summary}
We report an analysis of the CO J=2-1 and CS J=5-4 ALMA maps of the Flying 
Saucer, a nearly edge-on protoplanetary disk orbiting a T Tauri  
star located in the $\rho$ Oph molecular cloud. At the angular resolution of 
$0.5''$ (60 au at 120 pc) and in spite of some confusion in CO due to the 
background molecular clouds, we find that: 
\begin{itemize}\itemsep 0pt
\item the disk is in Keplerian orbit around a  0.57 $\Msun$ star
and nearly edge-on (inclined by $87^\circ$). 
It does not exhibit significant departure from symmetry, 
neither in CO nor in CS, 
\item direct evidence for a vertical temperature gradient
is demonstrated by the CO emission pattern. Quantitative estimates
are however limited by the spatial resolution.
\item disentangling between CO depletion and very low
temperatures is not possible because of the limited angular
resolution. Models with CO depletion in the mid-plane only
agrees marginally better, and the mid-plane temperature cannot
be significantly larger than 10 K at 100 au.
\item the CO emission is observed between 1 and 3 scale heights while 
the CS emission is located around one scale height. Sub-thermal
excitation of CS may explain this apparent difference.
\item CO is also observed beyond 
a radius of 230-260 au, in agreement with models predicting a
secondary increase of temperature due to higher UV flux penetration
in the outer disk. However, the limited angular resolution does not
rule out an alternate explanation with a molecular gap near $\rout$. 
\end{itemize} 
Finally, our results demonstrate that observing an edge-on disk is a 
powerful method to directly sample the vertical structure of 
protoplanetary disks  provided the angular resolution is high enough.
At least one data set with angular resolution around $0.1-0.2''$ is 
needed for a source located at 120-150 pc.

\begin{acknowledgements}
We thank the referee for constructive comments.  
This work was supported by ``Programme National de Physique Stellaire'' (PNPS
from INSU/CNRS.)
This research made use of the SIMBAD database,
operated at the CDS, Strasbourg, France.
This paper makes use of the following ALMA data:
   ADS/JAO.ALMA\#2013.1.00387.S. ALMA is a partnership of ESO (representing
   its member states), NSF (USA), and NINS (Japan), together with NRC
   (Canada),  NSC and ASIAA (Taiwan), and KASI (Republic of Korea) in
   cooperation with the Republic of Chile. The Joint ALMA Observatory is
   operated by ESO, AUI/NRAO, and NAOJ.
This paper is based on observations carried out with the IRAM 30-m telescope.
IRAM is supported by INSU/CNRS (France), MPG (Germany), and IGN (Spain).
VW's research is founded by the European Research Council (Starting Grant 3DICE, grant agreement 336474). 
\end{acknowledgements}


\bibliography{biblio-fls}
\bibliographystyle{aa}


\clearpage

\begin{appendix}
\label{app:rv}
\onecolumn

\section{PV diagram for an edge-on Keplerian disk}

\begin{figure}
    \centering
     \includegraphics[width=6.0cm]{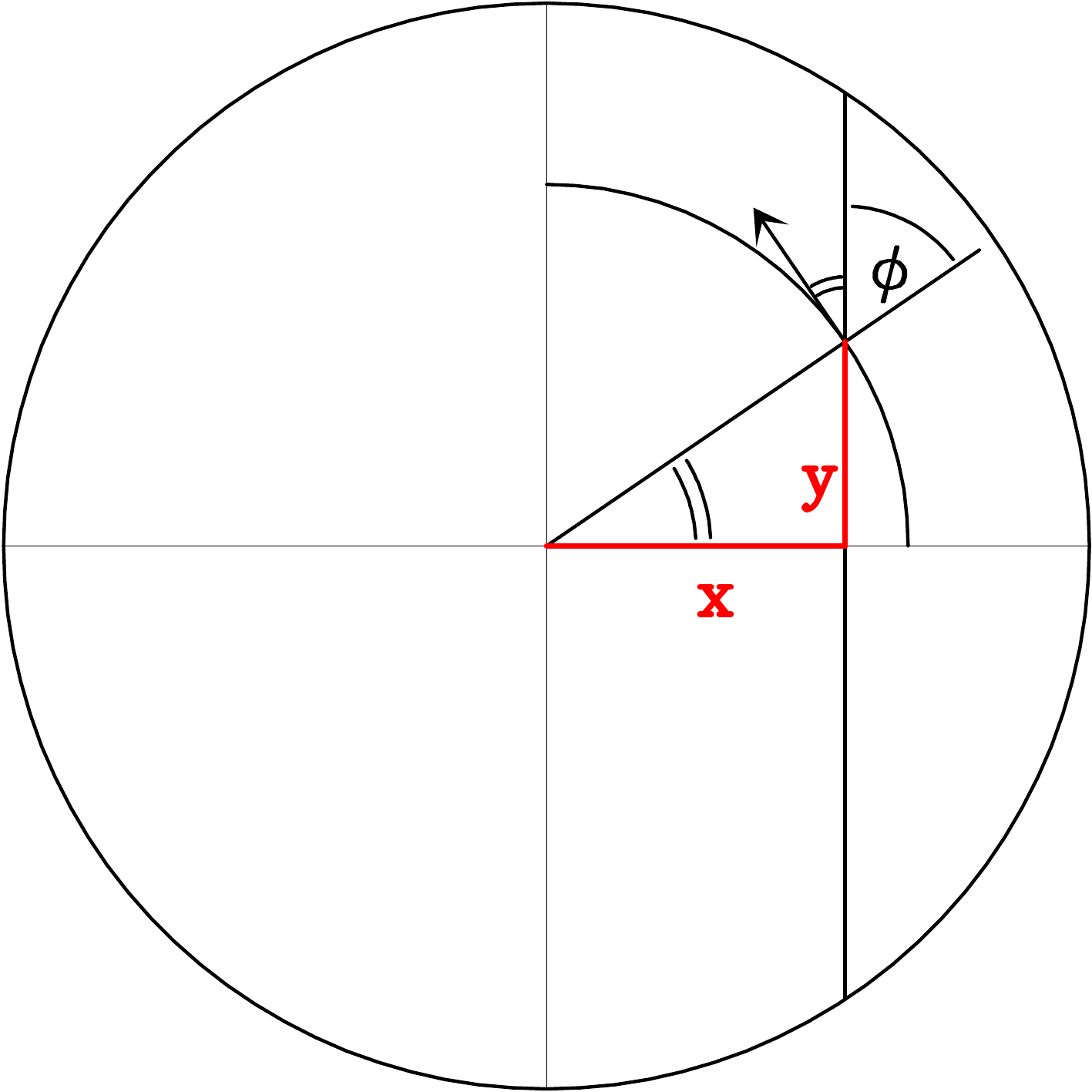}
    \caption{Definition of notations.}
    \label{fig:notations}
\end{figure}

\begin{figure}
    \centering
     \includegraphics[width=6.0cm]{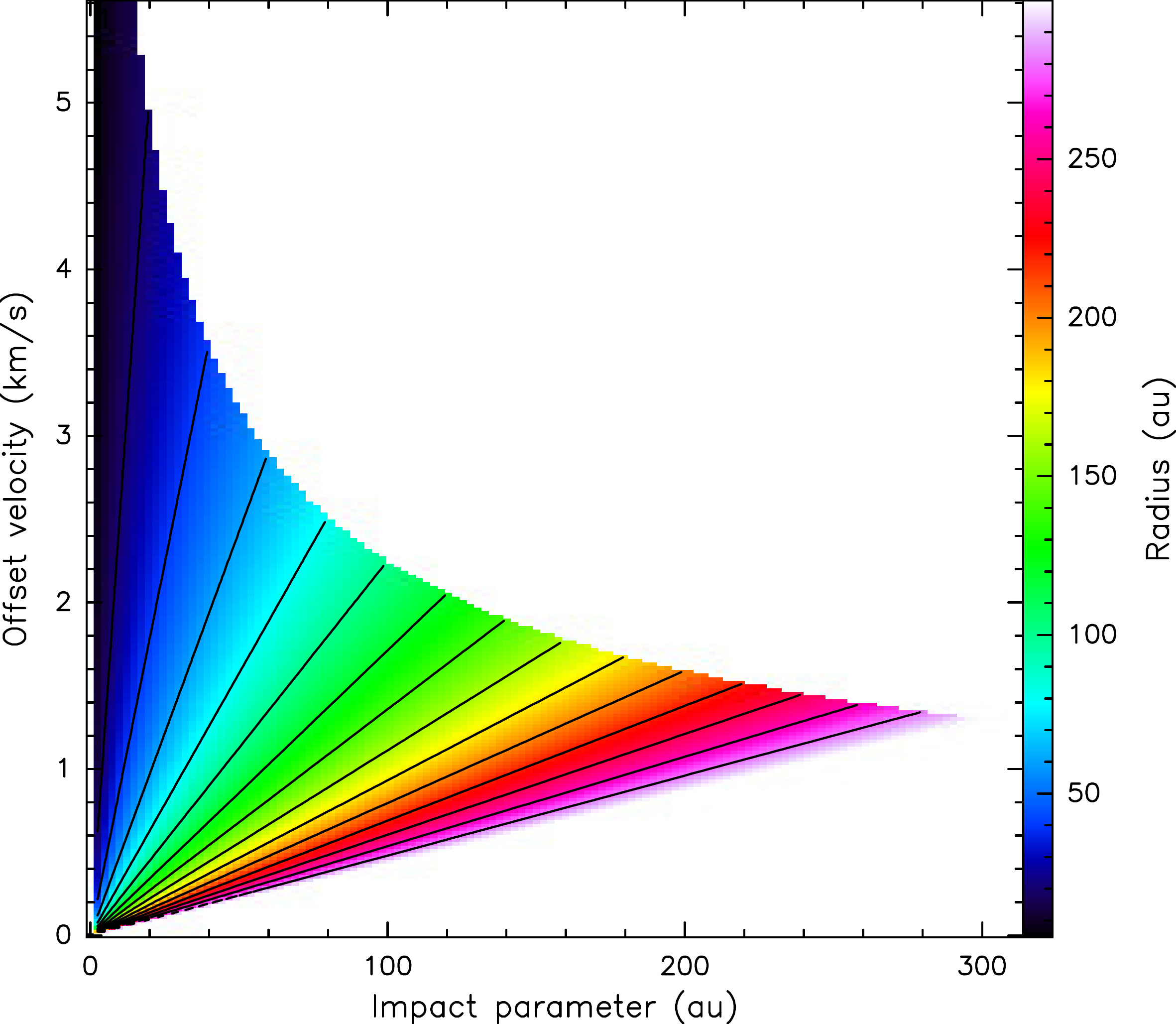}
    \caption{Radius as a function of Position and Velocity for
    an edge-on Keplerian disk}
    \label{fig:pv-b2}
\end{figure}

Let $x$ be the impact parameter in the disk, and $y$ the coordinate along the line
of sight, $r$ the radial distance
\begin{equation}
r(y) = \sqrt{x^2+y^2}
\end{equation}
where $y < y_m$, with $y_m$ given by
\begin{equation}
y_m = \sqrt{R_d^2-x^2}
\end{equation}
The angle $\phi$ is defined such that $x = r \sin{\phi}$ (see Fig.\ref{fig:notations})
The projected velocity along the line of sight is
\begin{equation}
V_y = \sqrt{GM/r} \sin{\phi} = \sqrt{GM/r} \frac{x}{r} = \sqrt{GM} \frac{x}{r^{3/2}}
\label{eq:vy}
\end{equation}
Thus we simply recover $r$
\begin{equation}
r = \left[ GM \left( \frac{x}{V_y} \right)^2 \right] ^{1/3}  = R_d \left( \frac{X}{V} \right)^{2/3}
\end{equation}
where $X= x/R_d$,$V=V_y/V_d$ and $V_d = \sqrt{GM/R_d}$ is the rotation velocity
at the outer disk radius. Since $x \leq r \leq R_d$, the above equation has a solution
provided $X < V < 1/\sqrt{X}$ (i.e. $V_y > V_d x/R_d$, and $V_y < \sqrt{GM/x}$).
So, for any given velocity $V_y$ and impact parameter $x$, we can solve for $r$, and
then recover \mbox{$y=\pm\sqrt{r^2-x^2}$} along the line of sight.

As a consequence, in the PV-diagram (showing functions of $(x,V)$) of a 
Keplerian disk, any line starting from ($x=0, V=V_\mathrm{sys}$) 
represent locii of constant radius. This is illustrated in 
Fig.\ref{fig:pv-b2}. We use this property
to directly recover the temperature as a function of radius 
(by taking the mean or the median for any given $r$) and altitude (by 
making cuts in the PV diagrams for different altitudes). 

For optically thick lines, this procedure yields the (beam 
averaged) excitation temperature, and thus the kinetic temperature if
the line is thermalized.  

For optically thin lines, the opacity is a function of $(r,x)$ 
because of the Keplerian shear which breaks the rotational symmetry. 
The above procedure thus yields a more complex function of the 
temperature and density, whose value however cannot exceed the 
excitation temperature at any radius $r$.

\section{Inclination effects}
\label{app:incli}

\begin{figure*}
    \centering
    \includegraphics[width=18.0cm]{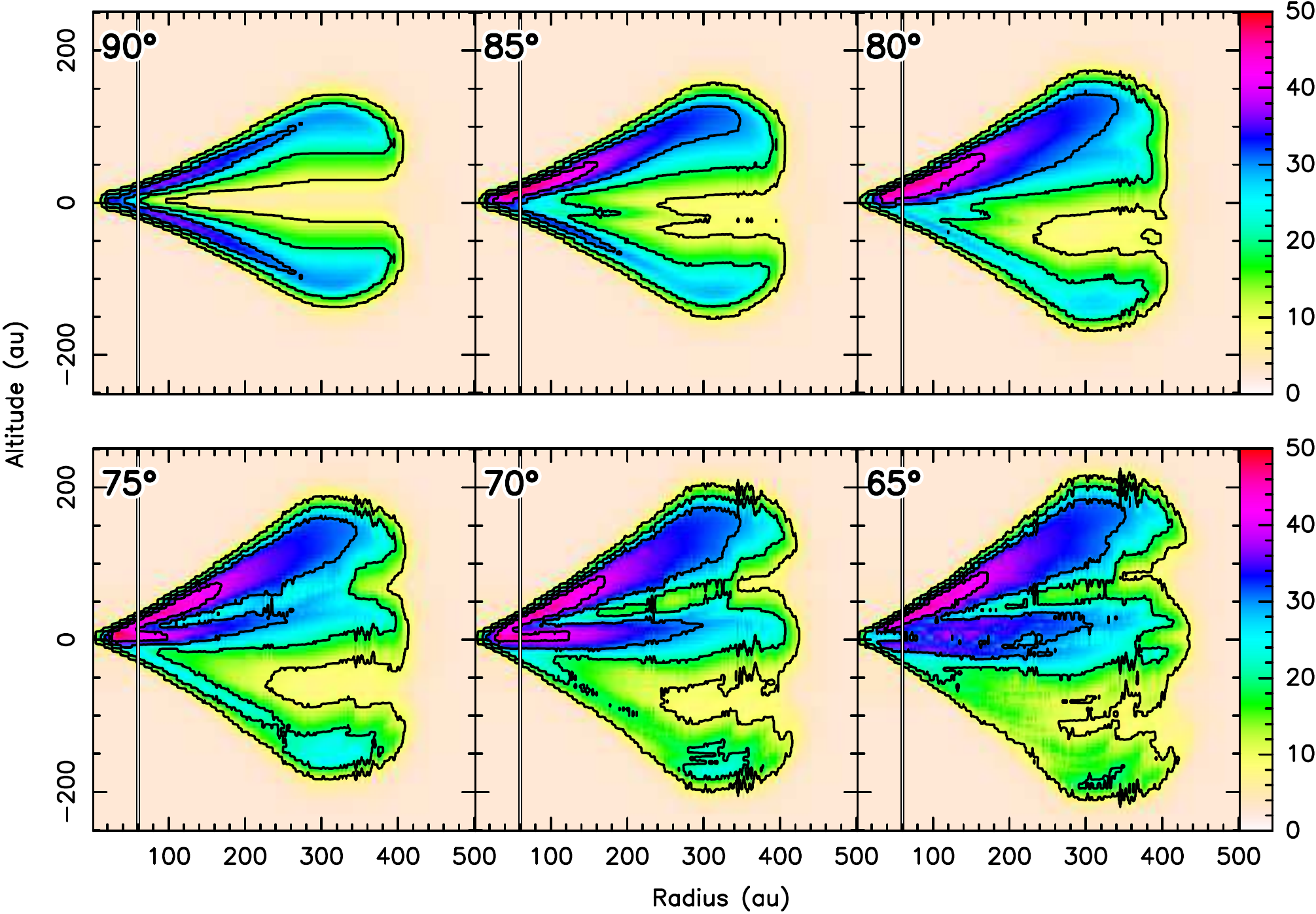}
   \caption{
    TRD  for CO J=2-1 (in Kelvin) of our
   best disk model for different inclinations. 
   Contours are in 10 K steps.
   }
  \label{fig:incli}
\end{figure*}

\textbf{
Fig.\ref{fig:incli} shows the expected TRDs of our fiducial disk model 
(Table 2) for different disk inclinations.  Since in this model, CO 
emits mostly 1 or 2 scale heights above the disk mid-plane, when the 
inclination differs from edge-on by more $h(r)/r$, the two opposite 
layers can project on the same side compared to the mid-plane 
projection. The farthest part of the disk projects to positive 
altitudes in Fig.\ref{fig:incli}. It appears warmer than the projection 
of the nearest part, because for the same impact parameter in altitude, 
the line-of-sight intercepts first warm  gas due to the disk flaring, 
while for the nearest part, the warm gas is hidden behind the forefront 
colder regions \citep[see][]{Dartois+etal_2003}. The depletion in the 
mid-plane make a clear distinction between the two emitting cones in 
the disk, resulting in a bright double layer at positive altitudes 
for $i=70-75^\circ$. At lower inclinations, the projected velocity 
gradient becomes insufficient to clearly separate these two layers on 
the TRD. For $i=80-85^\circ$, the two layers no longer project on the 
same side: a small lukewarm ``finger'' of emission appears at an 
altitude around $-20$ au.
}

\end{appendix}

\end{document}